\documentclass{ws-ijmpa}
\usepackage[super,nospace,compress]{cite}
\usepackage{graphicx}
\usepackage{subfigure}
\usepackage{hyperref}
\usepackage{xspace}
\usepackage{xcolor}

\newcommand*{\sqs}{\ensuremath{\sqrt{s}}\xspace}

\newcommand*{\Nch}{\ensuremath{N_\mathrm{ch}}\xspace}

\newcommand*{\RT}{\ensuremath{R_\mathrm{T}}\xspace}
\newcommand*{\RNC}{\ensuremath{R_\mathrm{NC}}\xspace}
\newcommand*{\NT}{\ensuremath{N_\mathrm{trans}}\xspace}
\newcommand*{\NNC}{\ensuremath{N_\mathrm{cone}}\xspace}
\newcommand*{\SO}{\ensuremath{S_\mathrm{0}}\xspace}
\newcommand*{\flt}{\ensuremath{\rho}\xspace}
\newcommand*{\pT}{\ensuremath{p_\mathrm{T}}\xspace}
\newcommand*{\pTtrg}{\ensuremath{p_\mathrm{T}^{\rm leading}}\xspace}

\newcommand*{\GeVc}{\ensuremath{\mathrm{GeV}/c}\xspace}

\newcommand*{\LcToDz}{\ensuremath{{\Lambda_{\rm c}^+}/{\rm D^0}}\xspace}
\newcommand*{\XicToDz}{\ensuremath{\ensuremath{\Xi_{\rm c}^{0,+}}/{\rm D^0}}\xspace}
\newcommand*{\Bp}{\ensuremath{\rm B^+}\xspace}
\newcommand*{\Bz}{\ensuremath{\rm B^0}\xspace}

\newcommand*{\LbToBp}{\ensuremath{{\Lambda_{\rm b}^0}/{\rm B^+}}\xspace}
\newcommand*{\LbToBpBz}{\ensuremath{{\Lambda_{\rm b}^0}/{(\rm B^+\mathord{+}B^0)}}\xspace}

\begin{document}
\markboth{L. V. Földvári, Z. Varga, R. Vértesi}{Event-activity-dependent beauty hadron enhancement}

%
\catchline{}{}{}{}{}
%

\title{Event-activity-dependent beauty-baryon enhancement in simulations with color junctions}

\author{Lea Virág Földvári}
\address{HUN-REN Wigner Research Centre for Physics,\\
Konkoly-Thege Miklós út 29-33, Budapest, 1121 Hungary;\\
Faculty of Science, Eötvös Loránd University,\\
Pázmány Péter sétány 1/A, Budapest, 1111 Hungary\\
foldvari.lea.virag@wigner.hu}

\author{Zoltán Varga\footnote{
Corresponding author.}}
\address{Wright Laboratory, Department of Physics, Yale University\\
New Haven, Connecticut 06520, USA\\
zoltan.varga@yale.edu}

\author{Róbert Vértesi}
\address{HUN-REN Wigner Research Centre for Physics,\\
Konkoly-Thege Miklós út 29-33, Budapest, 1121 Hungary\\
vertesi.robert@wigner.hu}

\maketitle

\begin{history}
\end{history}

\begin{abstract}
Recent results from ALICE and CMS show a low-transverse-momentum enhancement of charm baryon-to-meson production ratios over model predictions based on e$^+$e$^-$ collisions. Several mechanisms are proposed to understand this phenomenon. New measurements by the LHCb and ALICE experiments show a similar enhancement in the beauty sector. We explore this enhancement in terms of event activity using the color-reconnection beyond leading order approximation model. We propose sensitive probes relying on the event shape that will allow for the differentiation between the proposed beauty-production scenarios using freshly collected LHC Run-3 data, and we also compare these to predictions for charm. Our results will contribute to a deeper theoretical understanding of the heavy-flavor baryon enhancement and its relation to baryon enhancement in general. 
\end{abstract}

\keywords{baryon enhancement; heavy flavor; beauty production; event classification; color reconnection; underlying event; CERN LHC Coll.}
\ccode{PACS numbers: 13.30.Eg, 14.20.Mr, 13.25.Hw, 14.65.Fy, 14.40.Nd}

\section{Introduction}

Since its discovery by the RHIC experiments, collective flow has been considered a tell-tale sign of the strongly coupled quark-gluon plasma (QGP)\cite{Heinz:2013th}. Surprisingly, with the advent of the LHC, collective behavior was also observed in small collisional systems such as proton--proton\cite{CMS:2010ifv, ATLAS:2015hzw, CMS:2015fgy, CMS:2016fnw} (pp) or proton-lead\cite{CMS:2012qk, ALICE:2012eyl} (p--Pb). Similarly to collectivity, enhanced production of strange hadrons, which had long been anticipated as a signature of QGP formation\cite{Koch:1986ud}, was also found in small systems. Systematic studies have shown that strangeness enhancement depends on the event multiplicity, regardless of the colliding system~\cite{ALICE:2016fzo}. This leads to the question whether small droplets of quark-gluon plasma may come into being in collisions of small systems. Although there is no definite answer yet, evidence suggests that the observed collective phenomena can be explained by semi-soft vacuum-QCD effects, such as multiparton-interactions (MPI)\cite{Schlichting:2016xmj} with color-reconnection (CR)\cite{Ortiz:2016kpz} or the production of minijets\cite{Eskola:1997au} (semi-hard partons produced by incoming partons or bremsstrahlung).

Heavy-flavor hadroproduction in high-energy collisions is usually described based on the factorization approach, which states that the total cross section of the process is the convolution of three independent components: the parton distribution functions of the incoming nucleons (PDFs) or nuclei (nPDFs), the hard parton-parton scattering cross section, and the fragmentation function of heavy-flavor quarks into hadrons.
The fragmentation functions have traditionally considered universal across different collisional systems, supported by the different timescales of scattering and fragmentation processes, and a range of heavy-flavor meson measurements\cite{ALICE:2021mgk}. However, recent data from LHC experiments\cite{ALICE:2021rzj,CMS:2019uws,ALICE:2020wfu,LHCb:2018weo} show that models using fragmentation functions from e$^+$e$^-$ collisions significantly underestimate the charmed baryon-to-meson (\LcToDz and \XicToDz)%
\footnote{For the sake of simplicity, in this work we always account for charge conjugates without explicitly writing them out.}
ratios in pp collisions, thus questioning the universality of heavy-flavor fragmentation. The ALICE experiment also found that the charmed-baryon enhancement is multiplicity dependent in the mid-transverse-momentum regime\cite{ALICE:2021npz}, mirroring trends observed in strange-baryon production, where the yield of hyperons like $\Lambda$ and $\Xi$ increases with charged-particle multiplicity\cite{ALICE:2016fzo}. These observations point toward a common event-activity dependent source linked to collective phenomena, and suggest that certain features traditionally associated with the quark-gluon plasma formed in heavy-ion collisions might also be present in high-multiplicity pp collisions.

Several models attempt to explain the enhanced charm-baryon production based on quark coalescence\cite{Plumari:2017ntm,Song:2018tpv}, the existence of undiscovered excited charm-baryon states~\cite{He:2019tik}, or string formation with color reconnection beyond leading color approximation (CR-BLC)~\cite{Christiansen:2015yqa}.
Event activity observables such as final-state hadron multiplicity are strongly connected to MPI, and CR has been shown to produce collective behavior\cite{Ortiz:2016kpz}. The understanding of charmed-baryon enhancement in terms of event activity may therefore reveal the source of QGP-like behavior in collisions of small systems.
Further investigating the connection of charmed-baryon enhancement to the event activity, it had been found that the charm-baryon enhancement in the CR-BLC model is connected to the underlying event\cite{Varga:2021jzb,Varga:2023byp}.

Recent measurements of beauty baryon-to-meson ratios showed similar trends to those observed in charm\cite{LHCb:2019fns,LHCb:2023wbo}.
In the following we extend our study to the beauty baryon-to-meson production using simulations and compare our results to recent experimental data. We evaluate the event-activity dependence of the enhancement with different event classifiers to understand the origin of the beauty enhancement, and to lay ground for future LHC measuements that can pin down the source of QGP-like effects in pp collisions using heavy-flavor hadrons. We confirm that the recently introduced event-activity quantifier, flattenicity\cite{Ortiz:2022mfv}, works as an effective proxy for underlying-event activity.

\section{Analysis Method}

We used \textsc{Pythia} 8.309\cite{Bierlich:2022pfr} with soft-QCD settings to simulate pp collision events at $\sqrt{s}=13$ TeV.
The default \textsc{Pythia} 8 Monash tune\cite{Skands:2014pea} is optimized to describe minimum-bias, Drell--Yan and underlying-event data from the LHC, combined with data from SPS and the Tevatron LHC to constrain the energy scaling. However, it uses fragmentation functions based on e$^+$e$^-$ collisions, and fails to describe heavy-flavor baryon production in pp collisions. Fortunately, \textsc{Pythia} allows for the incorporation of additional features to describe observed data. The default \textsc{Pythia} CR mechanism, based on the MPI framework, can be replaced by a QCD-based scheme (CR-QCD) that minimizes string length and follows QCD color rules. A main feature of the new model is the introduction of junction structures. A particularly successful tune we utilized is CR-BLC mode 2, known for accurately reproducing \LcToDz ratios\cite{ALICE:2020wfu}. This model incorporates time dilation using the boost factor derived from the final-state dipole mass and mandates causal connections among all dipoles. We also used mode 0, which lacks time-dilation constraints and controls CR by the invariant mass scale parameter, and mode 3, which includes time dilation but requires only a single causal connection. Another feature we used is the thermodynamical string fragmentation\cite{Fischer:2016zzs}, where Gaussian suppression in mass and \pT is replaced by an exponential function, leading to different baryon-to-meson ratios compared to Lund fragmentation. Finally, we employed rope fragmentation with string shoving\cite{Bierlich:2020naj},
where strings close in spacetime are allowed to repel each other, potentially generating collective features in the final state.

We simulated 6 billion events with the CR-BLC mode 2, 7.3 billion events with the CR-QCD tune, and 0.5 billion events with the other settings.
We computed the \LbToBp and \LcToDz ratios as a function of \pT at mid-rapidity, $|y|<1$, in terms of several event-activity classifiers.
The heavy-flavor baryon-to-meson ratios are known to depend on the final-state multiplicity. In the first place we used the charged-hadron multiplicity \Nch in the central rapidity region $|\eta|<1$, as it is readily accessible with most collider experiments. We also required a track transverse momentum to be $\pT>0.15$ GeV/$c$ to replicate technical limitations\cite{ALICE:2014sbx}.

While \Nch is sensitive to particle production in the whole event including jets, MPI is mostly responsible for the generation of the underlying event\cite{Seymour:2013qka}. The underlying-event activity and the activity related to the leading hard process can be quantified separately by using the relative transverse event multiplicity quantifier \RT and the relative near-side jet-cone multiplicity quantifier \RNC, respectively. These are defined as
\begin{equation}
    \RT = \frac{\NT}{\langle \NT \rangle}\hspace{1 cm}\mathrm{and}\hspace{1 cm}\RNC= \frac{\NNC}{\langle \NNC \rangle}\ ,
\end{equation}
where \NT is the transverse charged-hadron multiplicity, defined as the number of charged hadrons with $\pT>0.15$ \GeVc in the range $\pi/2 < \Delta \varphi < 3\pi/2$, the
angle difference defined in the transverse plane with respect to the leading (highest-\pT) charged hadron. \NNC is the charged-hadron multiplicity within a cone with a radius of $r=\sqrt{(\Delta \varphi^2 + \Delta \eta ^2)}$, where $\Delta \varphi$ and $\Delta \eta$ are the relative azimuth angle and pseudorapidity compared to the leading hadron. This approach requires a well-identified hard process, therefore the leading hadron is required to have $\pTtrg > 5$ \GeVc, and consequently, the \RT and the \RNC is defined only in a fraction of events. 
To avoid the loss of statistics and characterize any event regardless of \pTtrg, other event classifiers can be utilized. One such quantity is the transverse spherocity\cite{Ortiz:2015ttf},
\begin{equation}
	\SO = \frac{\pi}{4} \min\limits_{\bf \hat{n}} \left( \frac{\sum_i \left| {\bf p}_{{\rm T},i} \times {\bf \hat{n}} \right| }{\sum_i p_{{\rm T},i}} \right)\ ,
\end{equation} 
where $i$ indexes the charged hadrons in the acceptance and ${\bf \hat{n}}$ is any unit vector in the azimuth plane. Pencil-like, "jetty" events that contain hard scattering are described by \SO $\rightarrow$ 1, while underlying-event-dominated, isotropic events are \SO $\rightarrow$ 0. However, the transverse spherocity concentrates on the central pseudorapidity, and it is not sensitive to the peripheries. To overcome this limitation, flattenicity (\flt) has been introduced\cite{Ortiz:2022zqr}. The $\varphi-\eta$ plane is split up into roughly squarish cells of equal area by dividing the $\eta$ axis into 10 and the $\varphi$ axis into 8 ranges. \flt is defined as the relative standard deviation of the average momentum in a cell,%
\footnote{Note that we use the definition based on particle \pT\cite{Ortiz:2022zqr} for compatibility with earlier studies\cite{Varga:2021jzb,Varga:2023byp}. A somewhat different definition based on particle yields is more suitable for detectors such as ALICE without forward tracking capabilities\cite{Ortiz:2022mfv}.}
\begin{equation}
\rho = \frac{\sigma _{p_{\rm T}^{\rm cell}}}{\langle p_{\rm T}^{\rm cell} \rangle}.
\end{equation}
Flattenicity can select hedgehog-like events without a characteristic jetty structure in high-multiplicity pp collisions. Since information is used from a broad rapidity range, any biases from the influence of the hard processes is mitigated\cite{Ortiz:2022zqr}.
All event characterization variables are divided into five classes, each containing a similar number of events, as shown in Tab.~\ref{tab:classes}.

\begin{table}[ht]
\centering
\tbl{Classification of events by activity.}{
\begin{tabular}{cccccc}
\toprule
class   & I      & II       & III      & IV       & V        \\ 
\colrule
\Nch  & $\leq 15$  & 16-30     & 31-40     & 41-50     & $\geq 51$  \\
\RT   & $\leq 0.5$ & 0.5-1     & 1-1.5     & 1.5-2     & $\geq 2$   \\
\RNC  & $\leq 0.5$ & 0.5-1     & 1-1.5     & 1.5-2     & $\geq 2$   \\
\SO   & 0-0.25     & 0.25-0.45 & 0.45-0.55 & 0.55-0.75 & 0.75-1     \\
\flt  & $\geq 2.5$ & 2-2.5     & 1.5-2     & 1-1.5     & 0-1 \\
\botrule
\end{tabular}}
\label{tab:classes}
\end{table}

\section{Results}

\begin{figure}[!h]
\centering
\includegraphics[width=.8\textwidth]{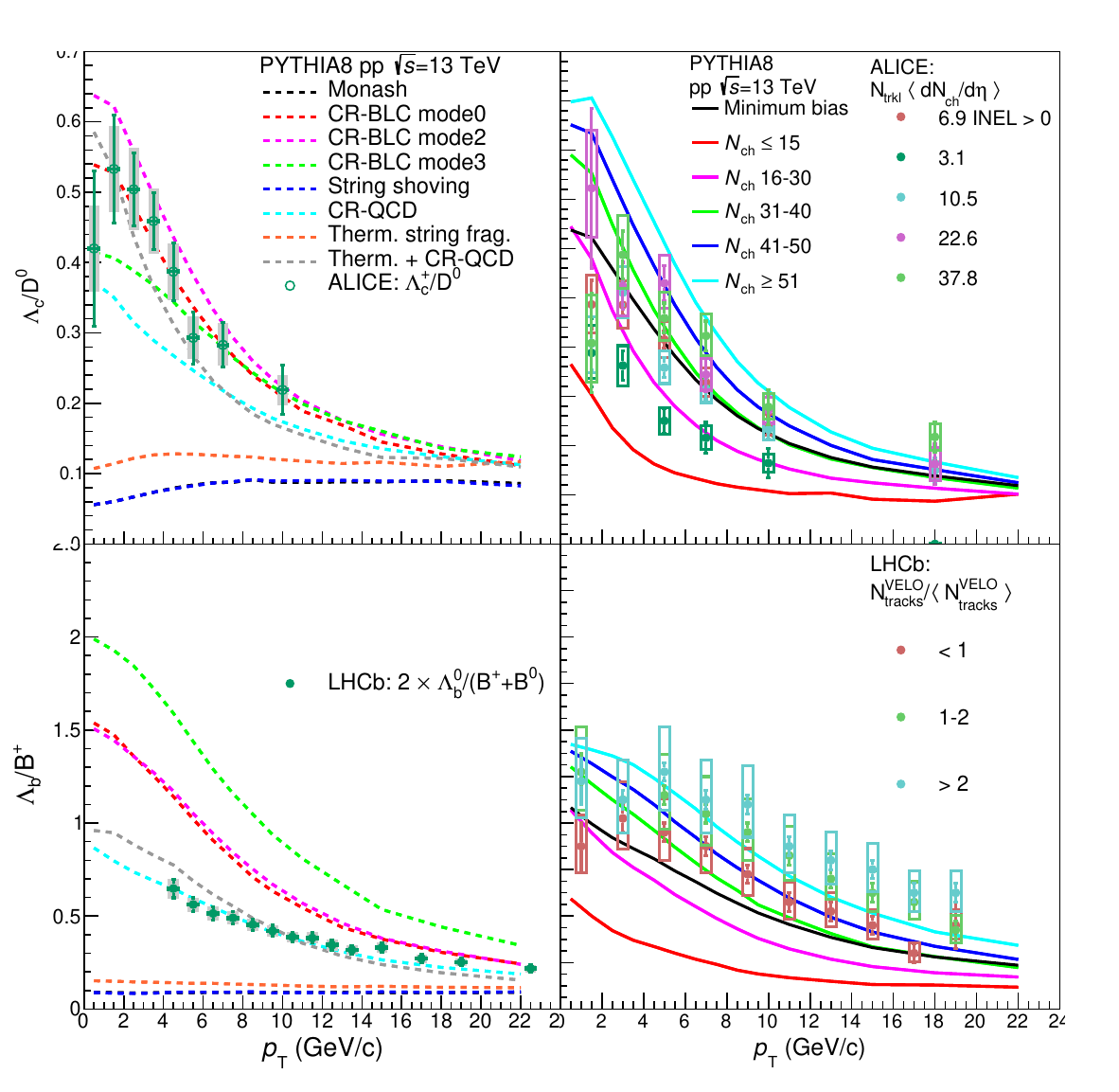}
\caption{Mid-rapidity ($|y|<0.5$) \LcToDz and \LbToBp ratios as a function of \pT in pp collisions at \sqs = 13 TeV from \textsc{Pythia} 8 simulations, compared to data. Panels from left to right, top to bottom: \LcToDz for several model settings\cite{Skands:2014pea,Christiansen:2015yqa,Fischer:2016zzs,Bierlich:2020naj}, compared to data from ALICE\cite{ALICE:2022exq}; \LcToDz from CR-BLC mode 2 in different \Nch classes, compared to ALICE multiplicity-dependent measurements\cite{ALICE:2021npz}; \LbToBp from CR-QCD for the above model settings, compared to LHCb forward-rapidity \LbToBpBz measurements\cite{LHCb:2019fns}; \LbToBp in different \Nch classes, compared to LHCb multiplicity-dependent measurements\cite{LHCb:2023wbo}.
}

\label{fig:multiplicity}
\end{figure}

The top and bottom left panels of Fig.~\ref{fig:multiplicity} show the \LbToBp and \LcToDz ratios at mid-rapidity as a function of \pT for different combinations of \textsc{Pythia} settings\cite{Skands:2014pea,Christiansen:2015yqa,Fischer:2016zzs,Bierlich:2020naj}, compared to \LcToDz results from ALICE\cite{ALICE:2022exq} at mid-rapidity, and LHCb measurements of \LbToBpBz taken at forward pseudorapidity ($2<|\eta|<5$)\cite{LHCb:2019fns}, respectively.
(Note that in the CR-BLC model class, the enhancement depends very weakly on the rapidity range.)%
\footnote{Assuming similar \Bp and \Bz production cross-sections, we scaled up the LHCb points with a factor of two.}

While all the models that incorporate the QCD-based color reconnection, including the CR-BLC modes, reproduce the heavy-flavor baryon enhancement trends, string shoving and thermodynamical string fragmentation alone do not make a qualitative difference. It is also to be recognized that while \LcToDz is quantitatively described by CR-BLC (particularly mode 2), the extent of \LbToBp enhancement is reproduced by CR-QCD without further adjustments. In the low-\pT range the CR-BLC simulations overestimate LHCb results by almost a factor of two, indicating that further model development is necessary to provide a general description on heavy-flavor baryon enhancement\cite{Altmann:2024odn}. 

The top and bottom right panels of Fig.~\ref{fig:multiplicity} show the  \LcToDz ratios simulated with CR-BLC mode 2, and \LbToBp ratios with CR-QCD, as a function of \pT in different \Nch classes, compared to multiplicity-dependent \LcToDz results from ALICE\cite{ALICE:2021npz} at mid-rapidity, and LHCb measurements of \LbToBpBz at forward pseudorapidity\cite{LHCb:2023wbo}, respectively.
As seen earlier for the case of charm\cite{Varga:2021jzb,Varga:2023byp}, both panels show stronger enhancement of the \LbToBp and \LcToDz ratios for higher-activity classes.

\begin{figure}[!h]
\centering
\subfigure{
\includegraphics[width=.4\textwidth,trim={10 20 20 50},clip]{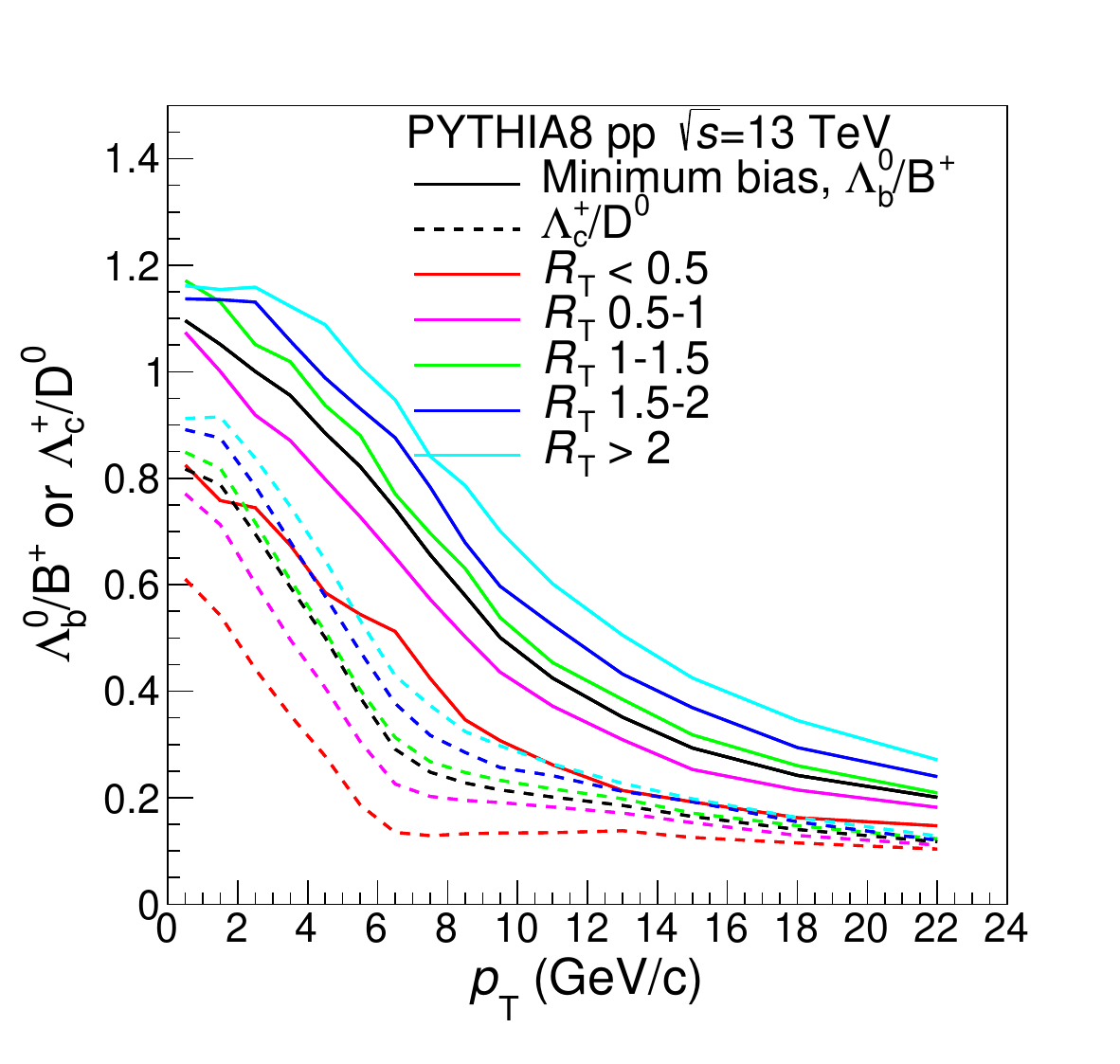}
\includegraphics[width=.4\textwidth,trim={10 20 20 50},clip]{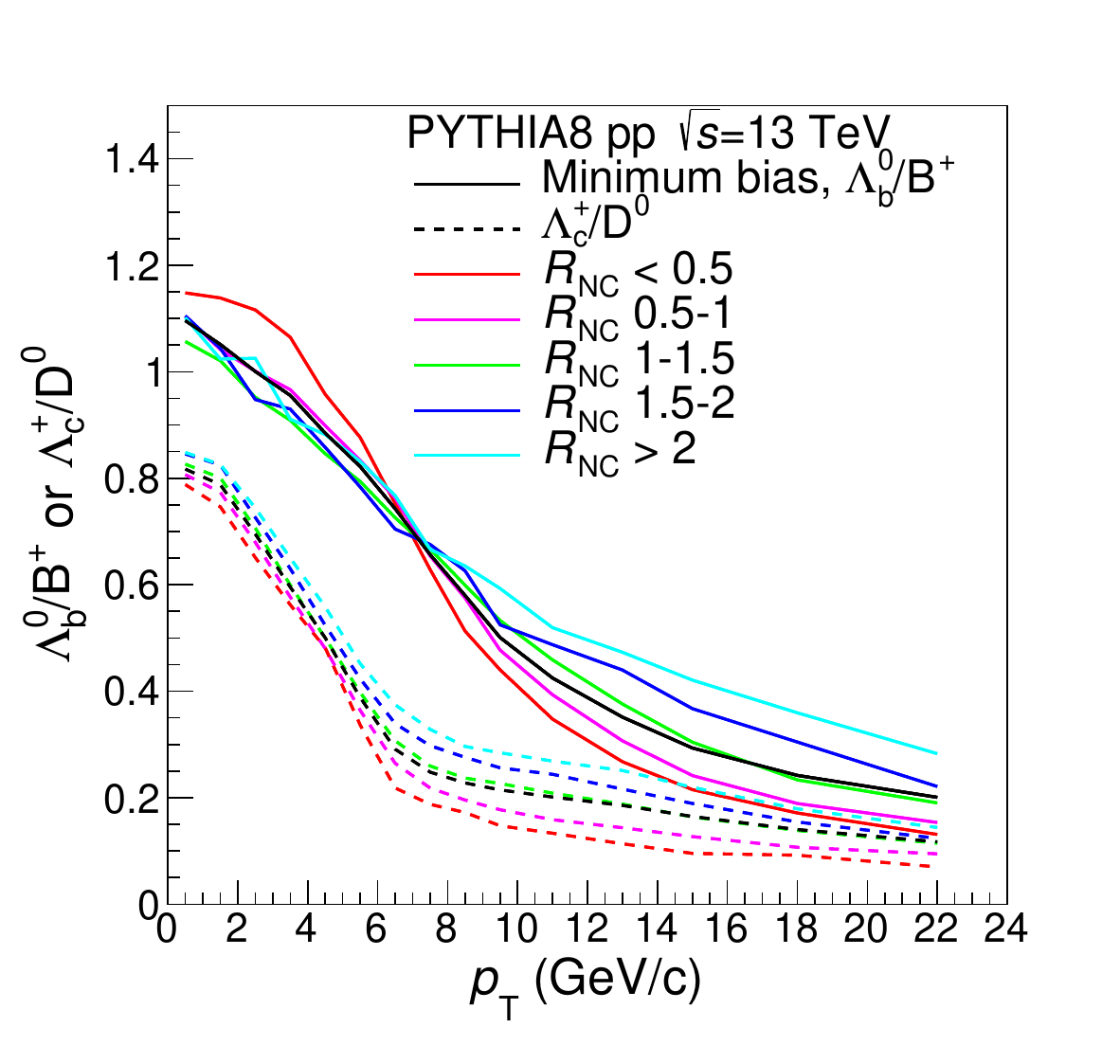}
}
\subfigure{
\includegraphics[width=.4\textwidth,trim={10 20 20 50},clip]{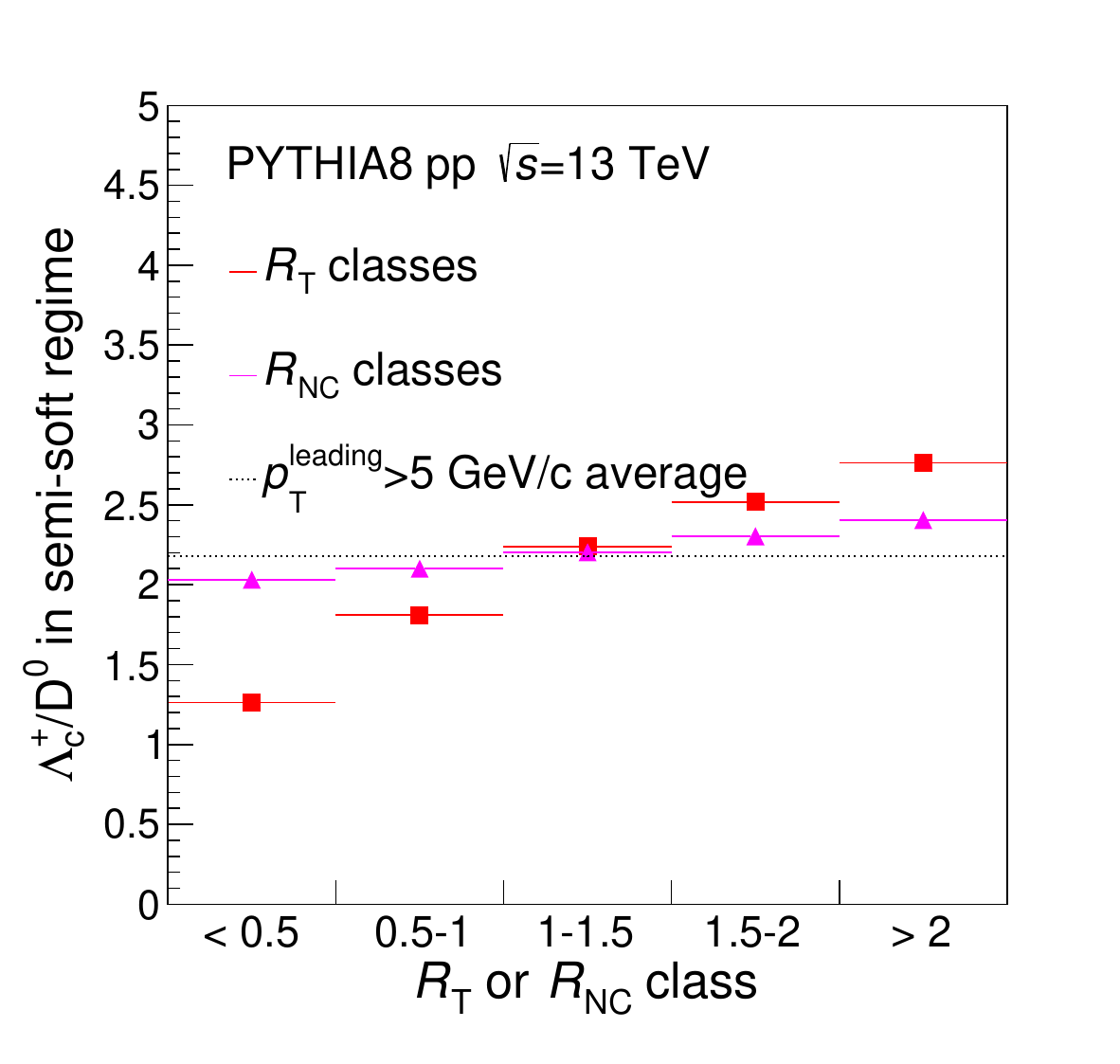}
\includegraphics[width=.4\textwidth,trim={10 20 20 50},clip]{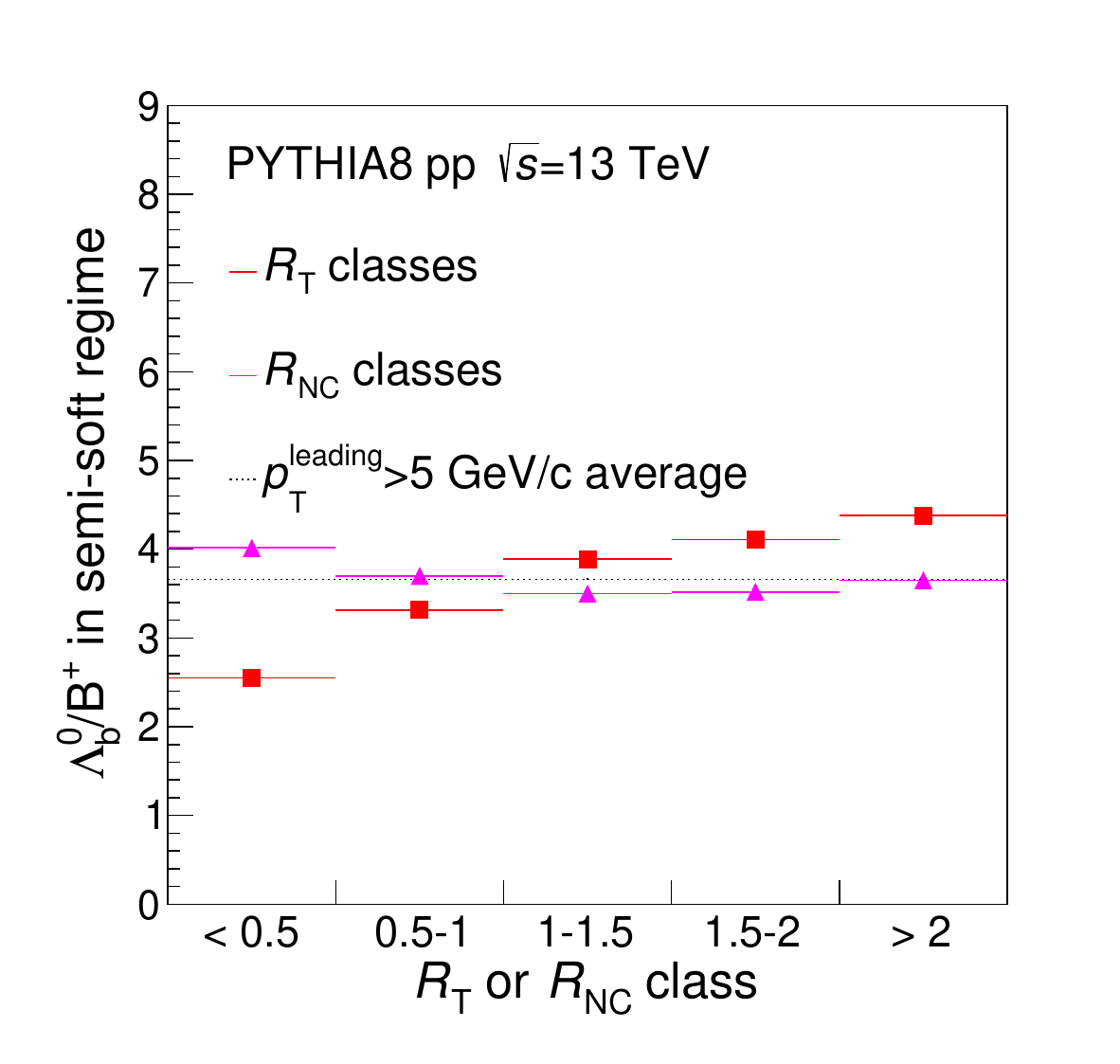}
}
\caption{\LcToDz and \LbToBp ratios from \textsc{Pythia} simulations. The top panels show the \LcToDz ratios from CR-BLC (dashed lines) and \LbToBp ratios from CR-QCD (solid lines) as a function of \pT in different \RT classes (left) and \RNC classes (right). The bottom left panel shows \LcToDz from, integrated over $2 < \pT < 6$ \GeVc as a function of \RT and \RNC bins, compared to the average (dotted line); the bottom right panel shows the same for the \LbToBp ratios.}
\label{fig:triggered}
\end{figure}

In the following, we show results with event-activity descriptors that are selectively sensitive either to the jetty or the soft part of the event. 
The top panels of Fig. \ref{fig:triggered} show the \LbToBp and \LcToDz ratios as a function of \pT for different \RT and \RNC classes, respectively. In both cases, the \pT-dependent trends in the \LbToBp ratios are similar to those in \LcToDz ratios \cite{Varga:2021jzb}. 
While the higher \RT classes show stronger enhancement, the enhancement is virtually independent of the \RNC classes below \pT $\approx$ 6 \GeVc. The bottom left panel shows \LcToDz integrated over $2 < \pT < 6$ \GeVc (the region relevant for semi-hard processes under investigation) as a function of \RT and \RNC bins, compared to the average, and the bottom right panel shows the same for the \LbToBp ratios. The results highlight that the enhancement increases with increasing underlying-event activity, while it does not depend on the activity in the jet region.

\begin{figure}[!h]
\centering
\subfigure{
\includegraphics[width=.4\textwidth,trim={10 20 20 50},clip]{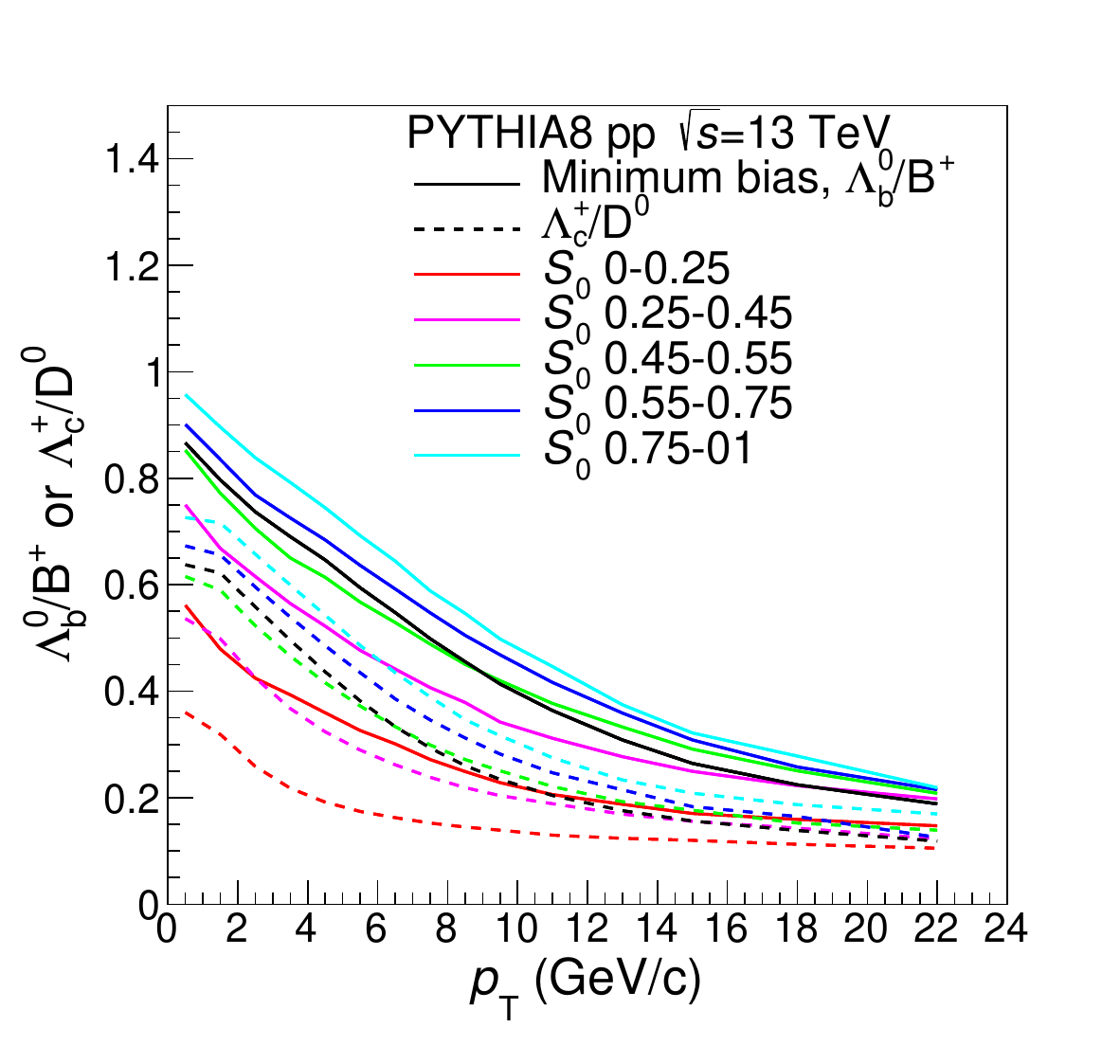}
\includegraphics[width=.4\textwidth,trim={10 20 20 50},clip]{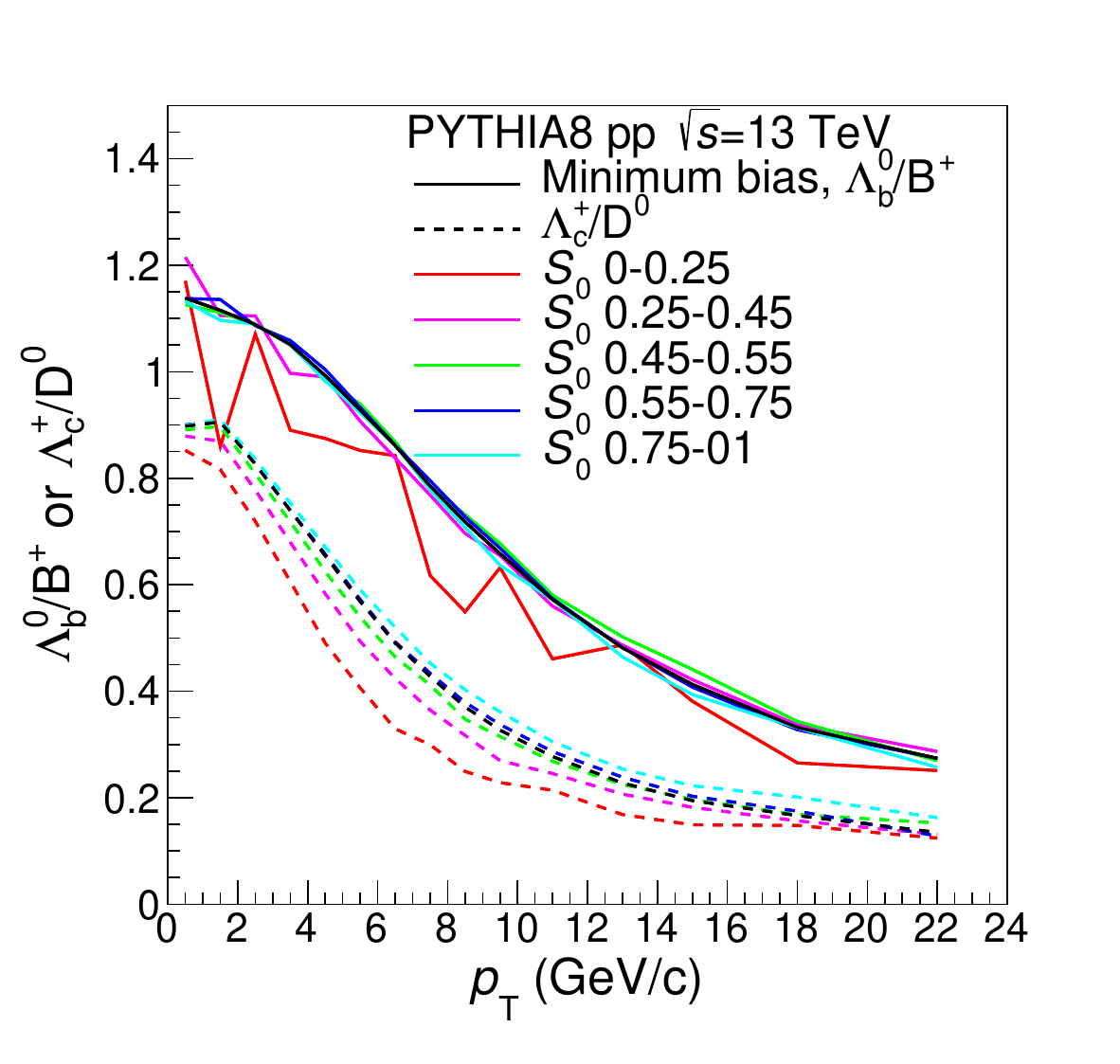}}
\subfigure{
\includegraphics[width=.4\textwidth,trim={10 20 20 50},clip]{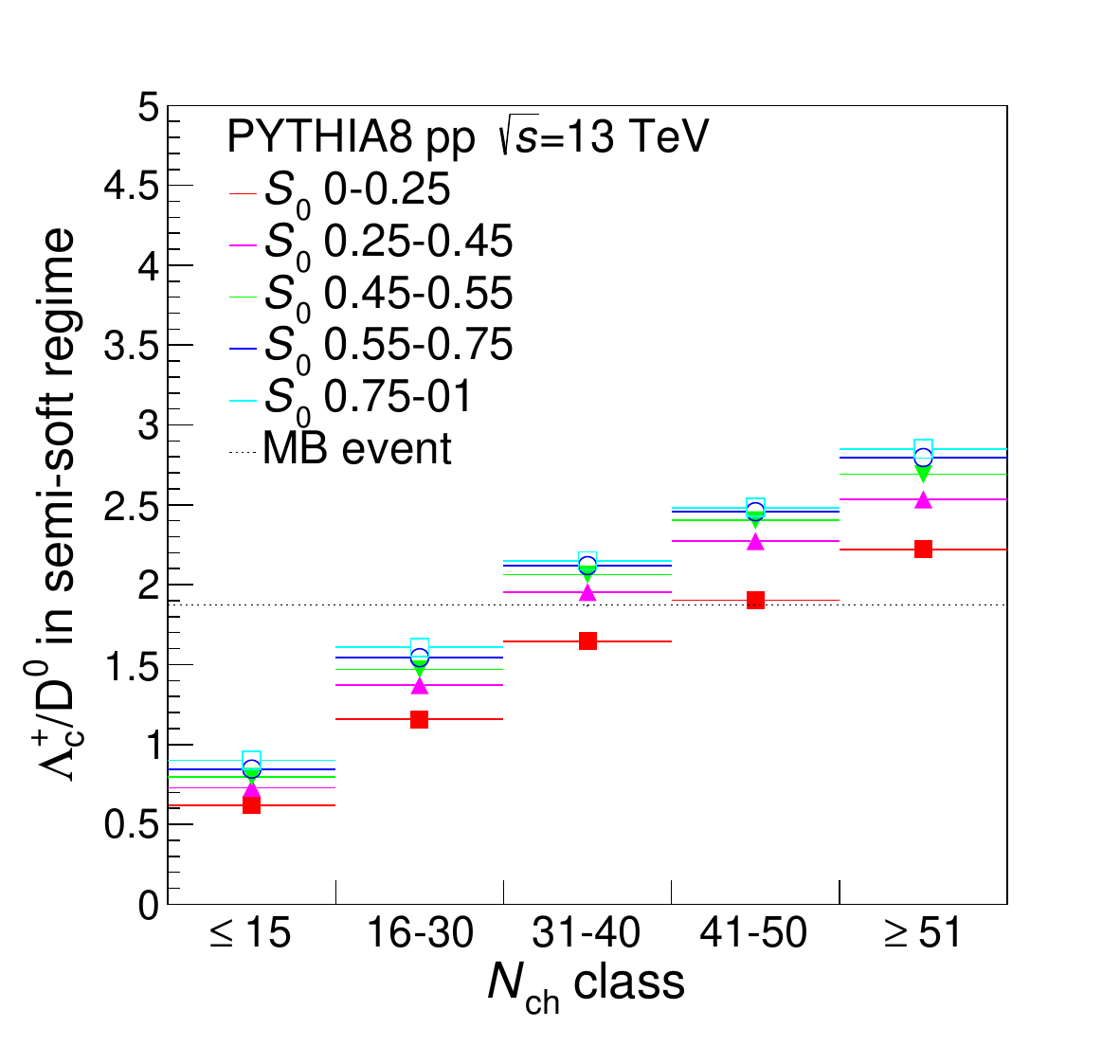}
\includegraphics[width=.4\textwidth,trim={10 20 20 50},clip]{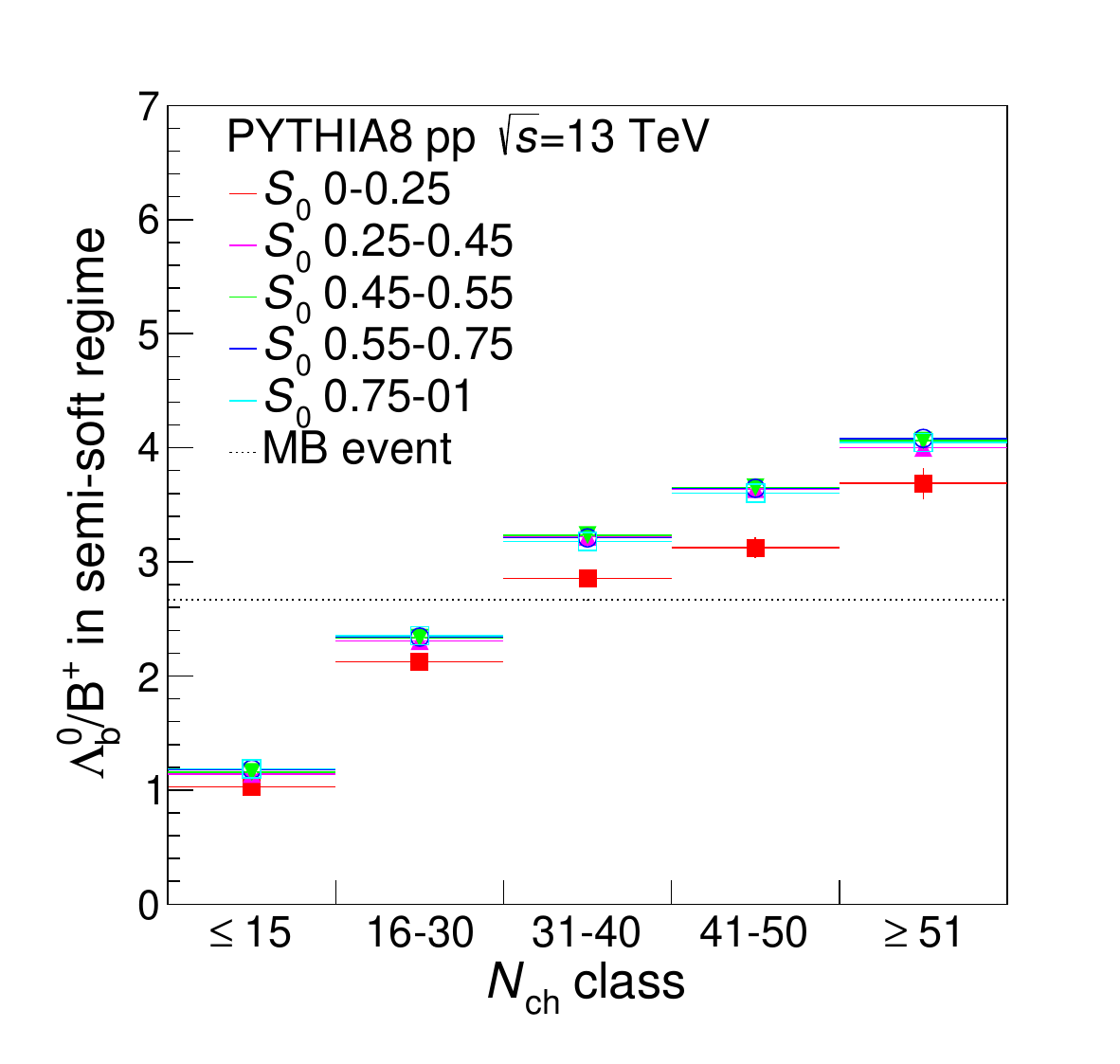}
}
\caption{\LcToDz and \LbToBp ratios from \textsc{Pythia} simulations. The top panels show the \LcToDz ratios from CR-BLC (dashed lines) and \LbToBp ratios from CR-QCD (solid lines) as a function of \pT in different \SO classes. The top left shows the ratios of the inclusive yield and the top right panel shows only the ratios for $\Nch > 50$. The bottom left panel shows \LcToDz ratios of the different \SO classes, integrated over $2 < \pT < 6$ \GeVc as a function of \Nch bins, compared to the average (dotted line), the bottom right panel shows the same for the \LbToBp ratios.}
\label{fig:spherocity}
\end{figure}

Fig. \ref{fig:spherocity} summarizes the spherocity-dependence of the heavy-flavor baryon-to-meson ratios from the simulations.
The top left panel shows the \LbToBp and \LcToDz ratios as a function of \pT in different transverse spherocity classes.
Since heavy-flavor production depends on different event-activity classifiers (in particular, \Nch and \SO) in a correlated way, in the top right panel we also show the results restricted to high-multiplicity events  ($\Nch > 50$). The bottom left and right panels show the \LcToDz and \LbToBp ratios, respectively, for different \SO classes integrated over $2 < \pT < 6$ \GeVc as a function of \Nch bins, compared to the average production (dotted line).
Like charmed hadrons beauty baryon enhancement in the higher-multiplicity bins slightly depends on \SO. However, this dependence is significantly weaker for beauty, making transverse spherocity an insensitive measure of underlying-event-dependent beauty hadron production.

\begin{figure}[!h]
\centering
\subfigure{
\includegraphics[width=.4\textwidth,trim={10 20 20 50},clip]{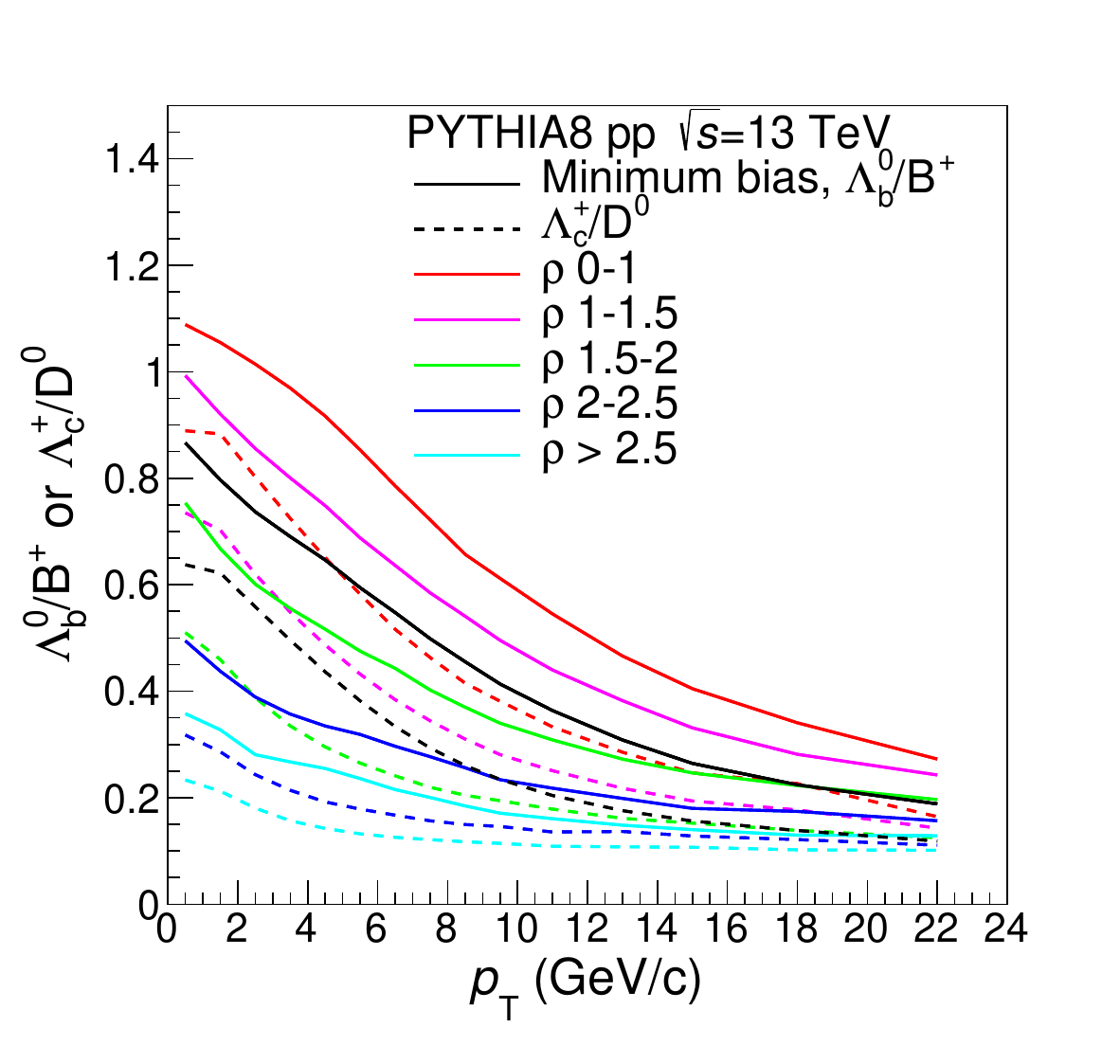}
\includegraphics[width=.4\textwidth,trim={10 20 20 50},clip]{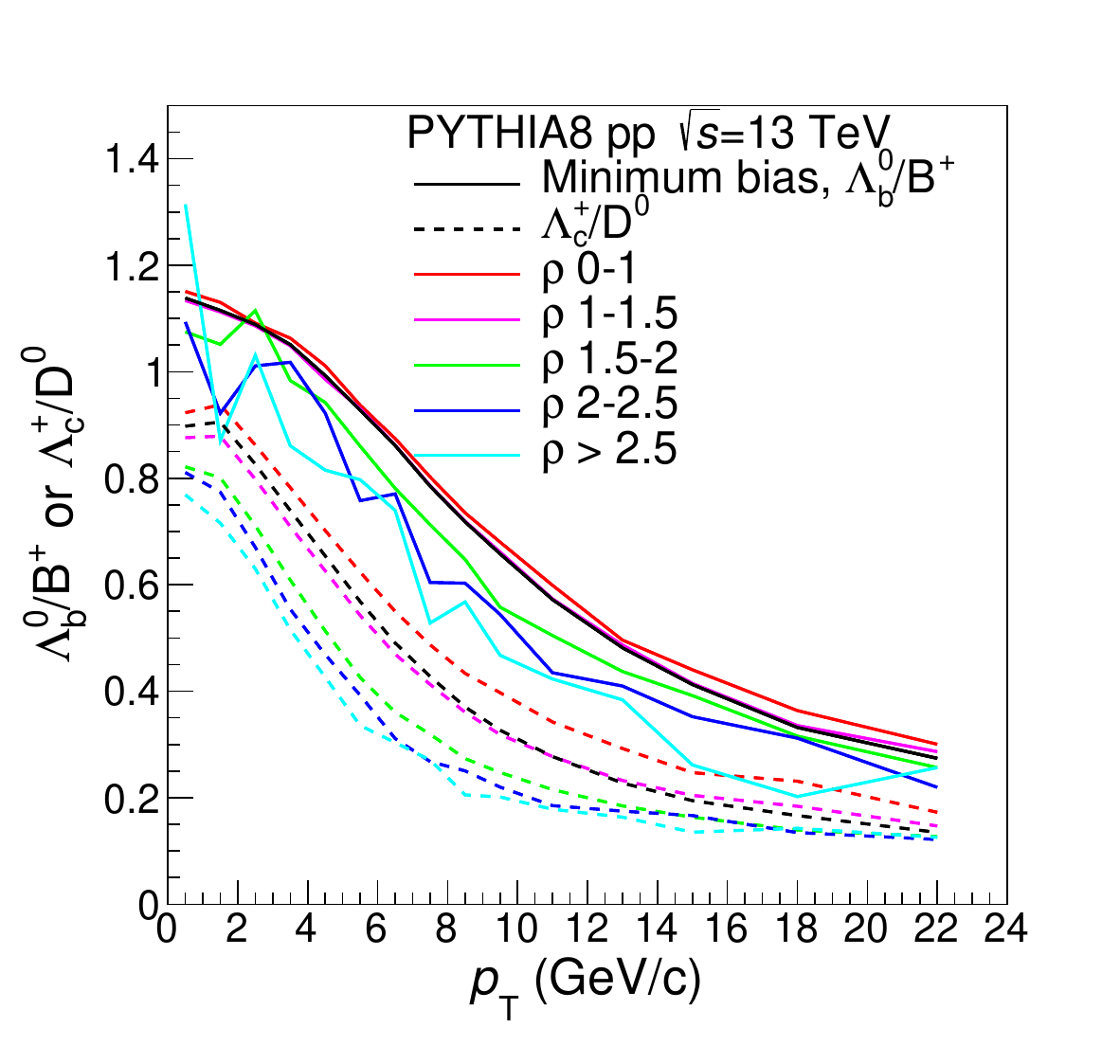}
}
\subfigure{
\includegraphics[width=.4\textwidth,trim={10 20 20 50},clip]{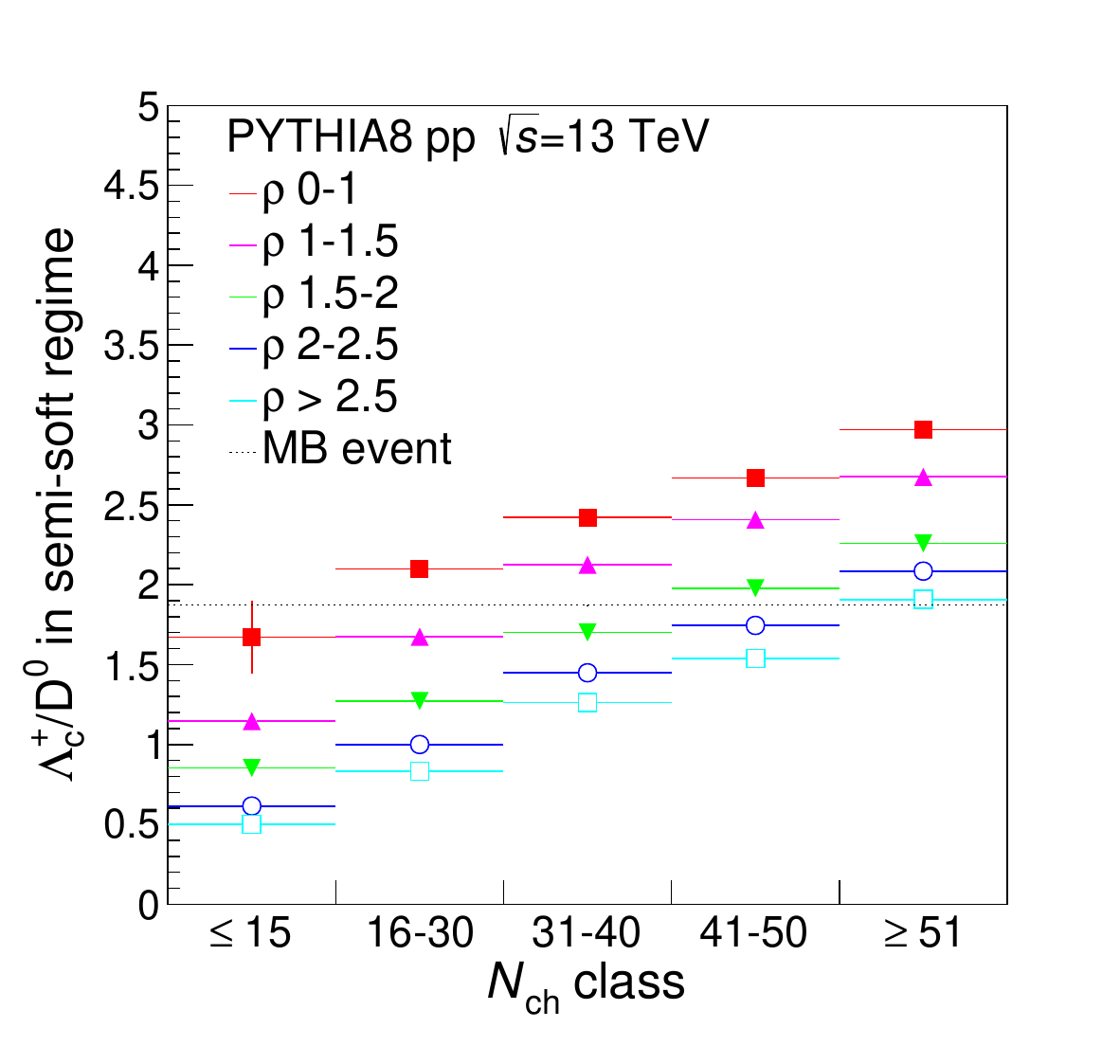}
\includegraphics[width=.4\textwidth,trim={10 20 20 50},clip]{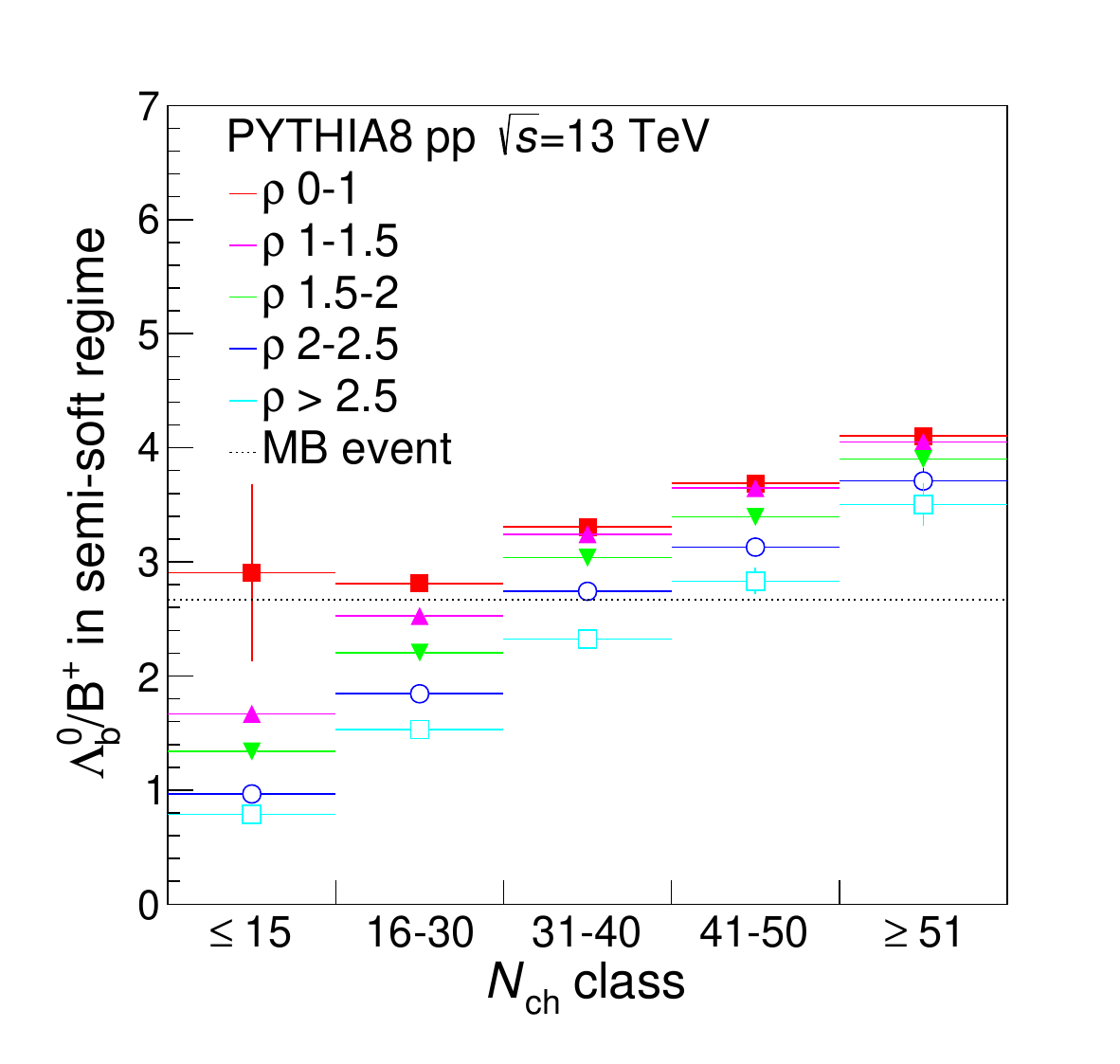}
}
\caption{\LcToDz and \LbToBp ratios from \textsc{Pythia} simulations. The top panels show the \LcToDz ratios from CR-BLC (dashed lines) and \LbToBp ratios from CR-QCD (solid lines) as a function of \pT in different \flt classes. The top left shows the ratios of the inclusive yield and the top right panel shows only the ratios for $\Nch > 50$. The bottom left panel shows \LcToDz ratios of the different $\rho$ classes, integrated over $2 < \pT < 6$ \GeVc as a function of \Nch bins, compared to the average (dotted line), the bottom right panel shows the same for the \LbToBp ratios.}
\label{fig:flattenicity}
\end{figure}

Fig. \ref{fig:flattenicity} outlines the \LbToBp and \LcToDz simulation results for the heavy-flavor baryon-to-meson ratios in terms of flattenicity.
The top left panel shows the \LbToBp and \LcToDz ratios as a function of \pT in different \flt categories.
Again, to avoid the correlation with \Nch, in the top right panel we also show the results restricted to high-multiplicity events ($\Nch > 50$). The bottom left and right panels show the \LcToDz and \LbToBp ratios, respectively, for different \flt classes integrated over $2 < \pT < 6$ \GeVc as a function of \Nch bins, compared to the average production (dotted line).
The curves representing different \flt values are distinctive, for both the charm and the beauty case, in the top right as well as the top left figure. Consequently, in the bottom figures, a clear separation with \flt can be seen in each \Nch class, without any remarkable dependence on the multiplicity. This indicates that flattenicity is an effective event classifier sensitive to the underlying event and the QCD processes, such as MPI, that drive it.

\section{Summary}

We compared the enhanced charm and beauty production in terms of several observables for event activity classification, using simulations that employ different tunes of the QCD-based color-reconnection scheme with color-string junctions. The charmed and the beauty baryon-to-meson production ratios show similar trends. Although a good quantitative description of the two heavy flavors with the same model settings is still missing, the demonstrated sensitivity on certain event-activity observables allowed us to assess their discrimination power in future measurements.

We explored the heavy-flavor baryon enhancement in terms of flattenicity, a new event activity classifier that is strongly related to the underlying event, and free from biases caused by mid-rapidity jet production. We see that, within the frames of the applied model class, it is plausible to assume that the heavy-flavor baryon enhancement stems from multiple-parton interactions, and it is connected to the underlying event.

By utilizing the methods that we outline, new high-luminosity LHC Run-3 data will be able to further constrain heavy-flavor fragmentation mechanisms, contribute to the development of more accurate models, and pin down the source of the observed baryon enhancement.

\section*{Acknowledgments}

This work was supported by the Hungarian NKFIH OTKA FK131979 and 2021-4.1.2-NEMZ\_KI-2024-00034 projects, as well as by the U.S. Department of Energy under grant number DE-SC0004168. The authors acknowledge the resources provided by the Wigner Scientific Computing Laboratory.

\bibliographystyle{ws-ijmpa}
\bibliography{references}
\end{document}